\documentclass[printer]{aa}  

\usepackage{graphicx}
\usepackage{txfonts}

\usepackage{color}           
\usepackage[normalem]{ulem}

\begin{document}

   \title{Interaction between the turbulent solar wind and a planetary magnetosphere: a 2D comet example}

   \subtitle{}

   \author{E. Behar
          \inst{1, 2}
          \and
          P. Henri\inst{2, 3}
          }

   \institute{Swedish Institute of Space Physics, Kiruna, Sweden\\
              \email{etienne.behar@irf.se}
         \and
             Lagrange, CNRS, OCA, Université Côte d'Azur, Nice, France
        \and
            LPC2E, CNRS, Univ. Orl\'eans, CNES, Orl\'eans, France
    }

   \date{Received September 15, 1996; accepted March 16, 1997}

 
  \abstract
   {Using the newly developed code \emph{Menura}, we present the first global picture of the interaction between a turbulent solar wind and a planetary obstacle in our solar system, namely a comet.}
   {This first publication aims at shedding lights on the macroscopic effect of the upstream solar wind turbulence on the induced magnetosphere of a comet.}
   {Using a hybrid Particle In Cell simulation code, we model a medium activity comet, using both a turbulent and a laminar solar wind input, for a direct comparison between the two regimes.}
   {We show how the turbulent characteristics of the solar wind lead to a smaller obstacle size. We then present how the upstream turbulent structures, traced by the perpendicular magnetic field fluctuations absent in the laminar case, self-consistently drape and pile-up around the denser inner coma, forming intense plasmoids downstream of the nucleus, pulling away dense cometary ion bubbles. This pseudo-periodic erosion phenomenon re-channels the global cometary ion escape and as a result, the innermost coma is found to be on average 45\% less dense in the turbulent case than predicted by simulating a laminar upstream flow.}
   {}


   \maketitle
%

\section{Introduction}

The solar wind -- a supersonic, radially expanding plasma escaping from the Sun -- can be described with two levels of complexity. First, by considering its average, background values, providing a global laminar picture in which most of the seminal studies on the heliosphere and planetary magnetospheres were founded \citep{biermann1951, alfven1957theory, parker1958apj, dungeyprl1961}. A second level of complexity introduces the turbulent nature of the flow, a combination of chaotic fluctuations within the magnetic field and particles density and velocity, adding up to their background, average values \citep{bruno2005lrsp}. In this phenomenology, energy (magnetic and kinetic) cascades from large to small scales, much like described by its neutral fluid analogue \citep{kolmogorov1941}, corresponding to a spectrum of fluctuations ranging over several decades of temporal and spatial scales. Eventually, this cascade leads to the dissipation of the energy at the smallest scales involved. Turbulence is suggested to play a key role in the acceleration of the solar wind, providing a continuous heating of the plasma from the solar corona and beyond \citep{cranmerptrs2015}.

When it comes to the formation and dynamics of planetary magnetospheres, the overwhelming majority of our knowledge was built on the laminar description of the solar wind. Specifically, all global numerical simulations of these interactions involve a steady, homogeneous plasma flow upstream of the obstacle \citep{schunk_nagy_2009, Ma2008, kallio2012}. A few recent exceptions can be pointed out, with for instance the Magneto-hydrodynamics (MHD) simulation using time-varying upstream conditions at the Earth \citep{lakka2017angeo}, two studies of a Coronal Mass Ejection interacting with the Earth, either using a MHD model \citep{lakka2019angeo} or a hybrid Particle-In-Cell (PIC) model \citep{moissard2022essoar}, and the hybrid PIC simulation of the effect of a pivoting magnetic field upstream of Mars \citep{romanelli2019grl}. 

However, a growing interest around the relationship between turbulence and magnetospheres is nowadays emerging. Aside the plethora of publications focused on turbulence within the Earth magnetosheath (\citet{rakhmanova2021fass} and references therein), several studies appeared on the topic of turbulence in the magnetospheres of outer planets and comets \citep{saur2021fass, ruhunusiri2020jgr}. Turbulence within the geomagnetic tail was also at the centre of many investigations, reviewed by \citet{antonova2021fass} and \citet{elalaoui2021fass}. More specifically, the effect of upstream turbulence on the terrestrial magnetosphere and its dynamics has also been investigated for decades, with results reviewed by \citet{damicis2020fass} and \citet{guio2021fass}. This sizable literature outlines one main lacking aspect: a numerical tool for the global simulation of these turbulent interactions. This is where the recently developed code \emph{Menura} \citep{behar2022angeo} positions itself, allowing the injection of a fully turbulent flow upstream of an obstacle.

This publication presents the first application of the code, and focuses on a cometary magnetosphere characterised by a neutral outgassing typical for a heliocentric distance of about two astronomical units (au). Since the dawn of solar and space physics, comets have been emblematic tracers of the solar wind, hinting at its very existence \citep{biermann1951, alfven1957theory}, while interplanetary sector boundaries as well as CMEs were notably analysed using remote observations based on the intermittent disconnections of comet tails \citep{niedner1978apj, vourlidas2007apj}. More recently, comets were used once more to trace some of the solar wind turbulent parameters \citep{deforest2015pj} as well as its speed \citep{cheng2022apj}. In addition to this historic role, a great amount of knowledge on cometary magnetospheres was produced during the last decade in the context of the European mission Rosetta \citep{goetz2022ssr}. This provides us with a solid background for a first global exploration of such a turbulent interaction.

Because of the multi-scale nature of turbulence, its numerical simulation is intrinsically expensive. For this first study, we made the choice of properly resolving a wide range of scales, from the magnetospheric to the ion scales, below their inertial length. To allow a practical handling of this problem, and specifically to quickly iterate the simulation and its analysis, we chose to work in a two-dimensional spatial domain, with velocities and field components described in a three dimensional space. This inherently limits the significance of our findings. To this extent, the aims of this first study are not to bring definitive results, but to properly illustrate the capacity of this new numerical approach, to give a first example on how its products can be tackled, and most importantly to highlight new aspects of the interaction, to be later on verified by a three-dimensional approach, in the limit of realistic computing resources. It should however not be underestimated that the best resolution achieved by a 2D approach, and therefore the corresponding plasma mechanisms, cannot be verified by a 3D approach with an equal computational power. Therefore a 2-dimensional approach cannot be limited to a role of pathfinder: it may very well demonstrate mechanisms otherwise not reproducible.

This first publication focuses on the effect of turbulence on the obstacle itself, looking into scales around and higher than the ion scale, while the characteristics of the turbulent flow within the obstacle is left for future studies.

\section{The Model} \label{sec:model}

The numerical model used to investigate the interaction of the solar wind -- either laminar or turbulent -- with a planetary obstacle is described and tested in~\cite{behar2022angeo}. It is based on a hybrid Particle-In-Cell (PIC) model implementation of the Vlasov-Maxwell equations, including a source term for the distribution function: the ions are described as massive particles, the electrons are considered as massless and charge-neutralising, while the electromagnetic fields together with the particles' moments are gathered at the nodes of a regular grid. At the core of the model is a generalised Ohm's law that computes the electric field provided the magnetic field and the particles' moments. \emph{Menura} uses the following formulation:

\begin{equation}\label{eq:ohm}
    \mathbf{E} = -\mathbf{u_i}\times\mathbf{B} + \frac{1}{e\, n}  \mathbf{J} \times \mathbf{B} - \frac{1}{e\, n} \nabla\cdot p_e - \eta_h \nabla^2 \mathbf{J}
\end{equation}
    
with $\mathbf{u_i}$ the ion bulk velocity, $n$ the plasma density, $\mathbf{J}$ the charge current, $p_e$ the electron pressure. Additionally, a term of hyper-resistivity is used to dampen small scales numerical oscillations, with the coefficient $\eta_h$ multiplying the laplacian of the current. Through the Faraday's law, this corresponds to a diffusion term. $\eta_h$ is taken to be $2.5\, 10^{-4}$ in the entire study. The electron pressure is obtained assuming it results from a polytropic process, with an index of 1 used throughout the study, corresponding to an isothermal process.

The code uses normalised units, with distances expressed in units of the background proton inertial length $d_{i0}$ and time in units of the inverse of the background ion cyclotron frequency $\omega_{ci0}^{-1}$.

The model is based on a two-steps procedure. First, a turbulent flow is generated, in the absence of the obstacle, as further described in Section \ref{sec:turbulent}. Second, this turbulent solar wind is injected in a simulation domain containing an obstacle. Since Step 1 is solved in a fully periodic domain, the injection of Step 1 outputs within the domain of Step 2 can be done periodically as well (see \citet{behar2022angeo} for more details). For such a medium activity comet, the domain boundaries parallel to the flow are also kept periodic.

During both steps, the equations and all variables are solved and expressed in the solar wind reference frame. It is therefore the obstacle which is moving through the solar wind. The domain is kept centred on the obstacle by regular copies and shifts of the fields and the particles, as illustrated in \citet{behar2022angeo}. To that extent, the solar wind is not really \emph{injected} in the domain but \emph{laid down} in front of the moving object. Because the object reference frame is the one used in all planetary simulation codes we have encountered, the vocabulary will unavoidably present some ambiguities between the two frames. In the following sections, we describe results and mechanisms in either the solar wind reference frame or the object reference frame, making sure to specify which.

During Step 2, in order to simulate a comet, a collection of cometary ions is added at each time step, as described in the next section.

In order to properly appreciate the influence of the incoming turbulent flow on the interaction, the model can also be used to send a laminar flow on the obstacle, all other parameters kept equal. We first describe this laminar case in Section~\ref{sec:laminar}, before considering the turbulent case in Sections \ref{sec:turbulent} and \ref{sec:comp}.

\section{The Obstacle: a comet} \label{sec:laminar}

\begin{figure}
    \centering
    \includegraphics[width=\linewidth]{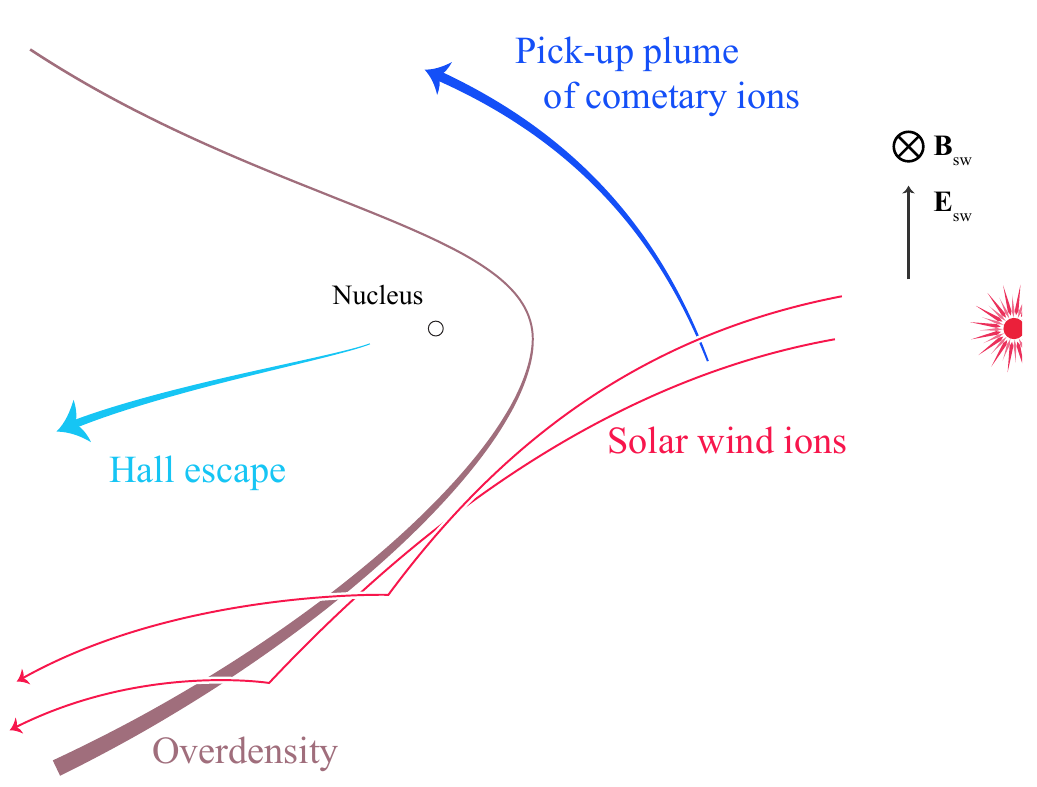}
    \caption{Schematic of a medium activity comet, highlighting the asymmetric dynamics of the solar wind ions, and indicating the two major escape channels for cometary ions.}
    \label{fig:schema}
\end{figure}

Without gravity, intrinsic magnetic field or solid central body (the size of the nucleus being negligible with respect to the dynamical scales of the system), and 
without ion collisions, the numerical modelling of such an obstacle is fairly straightforward: cometary ions are introduced at each iteration, according to a given ion production rate

\begin{equation}\label{eq:ionisation}
    q_i(r) = \nu_i \cdot n_0(r) = \frac{\nu_i Q}{4 \pi u_0 r^2} ,   
\end{equation}

with $r$ the distance from the comet nucleus, $\nu_i$ the ionisation rate of cometary neutral molecules, $n_0$ the neutral cometary density, $Q$ the neutral outgassing rate, $u_0$ the radial expansion speed of the escaping neutral cometary atmosphere. At the beginning of the run, the number $\alpha$ of particles to be added in the proximity of each grid node is first calculated using Eq. \ref{eq:ionisation} (together with the duration of one iteration). At each time step, we inject a constant number of particles with random positions within the cell surrounding each grid node, a number given by the integer part of $\alpha$. Each iteration, we produce and compare one more random number between 0 and 1 to the remaining part of $\alpha$, to decide whether one additional particle is created or not. In the sub-region surrounding the centre of the comet, where the distribution of neutral molecules shows the highest (radial) derivative, it is necessary to use a finer sub-grid to estimate these $\alpha$ values, and then average them over the main grid nodes closest to the nucleus, in order to not underestimate the local creation. We use a sub-grid ten times finer than the main grid, after having verified that no significant change using an even finer sub-grid can be found.

The highest cometary ion density is found close to the nucleus, a region within which noticeable plasma structures appear: this is the interaction region we simulate. At 2 au away from the Sun, these plasma structures and boundaries appear over the ion kinetic scales, characterised by the cometary ion gyro-radius, which will be discussed in further details in the Section~\ref{sec:comp}. This interaction region is sketched in Figure \ref{fig:schema}. Upstream of the nucleus, seldom newly born cometary ions are \emph{picked up} by the solar wind electric field, and start their cycloid motion. Closer to the nucleus, as their density is much higher, they form a noticeable density structure, which we hereafter refer to as the \emph{"pick-up plume channel"} (analog to the so-called pick-up plume at Mars \citep{dong2015grl}), the first significant escape channel for cometary ions. This early phase of the gyration is represented in Figure \ref{fig:schema}.

As the solar wind permeates the ionised cometary atmosphere, the total bulk velocity of the plasma decreases while the density of cometary ions increases. The frozen-in magnetic field \emph{piles up} on the denser coma, its amplitude increasing. Because the coma is denser close to the nucleus than elsewhere, the magnetic field additionally \emph{drapes} around the nucleus, in the iconic shape established by \citet{alfven1957theory}. Close to the nucleus, the magnetic field strength and its distortion become so intense that eventually, through the Hall component of the electric field (given by the local curl of the magnetic field under the Darwin approximation $\partial_t\mathbf{E}<<\mathbf{J}$), the dense inner coma is accelerated downstream, presenting a second escape channel for cometary ions, also indicated in Figure \ref{fig:schema} \citep{behar2018aa_b}. In the following, we refer to this cometary ion escape channel as the "\emph{Hall escape channel}". 

In the schematic, the cometary pick-up ions are accelerated upward, and as a result of momentum conservation, solar wind ions are deflected downward. This kinetic effect results in the formation of a solar wind over-density highly asymmetric further away from the Sun \citep{behar2018aa}, which transition to a more symmetric structure at smaller heliocentric distances \citep{hansen2007ssr}. Together with the structures formed by the pick-up plume and the Hall escape channels, this solar wind overdensity is the third main cometary ion structure generated by the solar wind-comet interaction at such a heliocentric distance. Each one of them will be tackled in the rest of this study to diagnose the effect of upstream solar wind turbulence on the plasma environment of the obstacle.

\section{The incoming flow: a turbulent solar wind} \label{sec:turbulent}

\begin{table}
    \centering
    \begin{tabular}{c|c}
        $B_0$ & 3.0 nT \\
        $n_0$ & 3.0 cm$^{-3}$ \\
        $\omega_{ci0}$ & 0.29 s \\
        $d_{i0}$ & 131 km \\
        $v_{A0}$ & 38 km/s \\
        $v_{th i0}$ & 38 km/s \\
        $\beta_{i0}=\beta_{e0}$ & 1 \\
        $\text{rms}(B_{\perp})/<B>$ & 0.32 \\
    \end{tabular}
    \caption{Characteristics of the turbulent solar wind, with $\text{rms}(B_{\perp}) = \sqrt{\text{rms}(B_{x}) + \text{rms}(B_{y})}$ and $<B>$ the average value of the total amplitude over the domain.}
    \label{tab:decay_param}
\end{table}
\begin{figure*}
    \includegraphics[width=\linewidth]{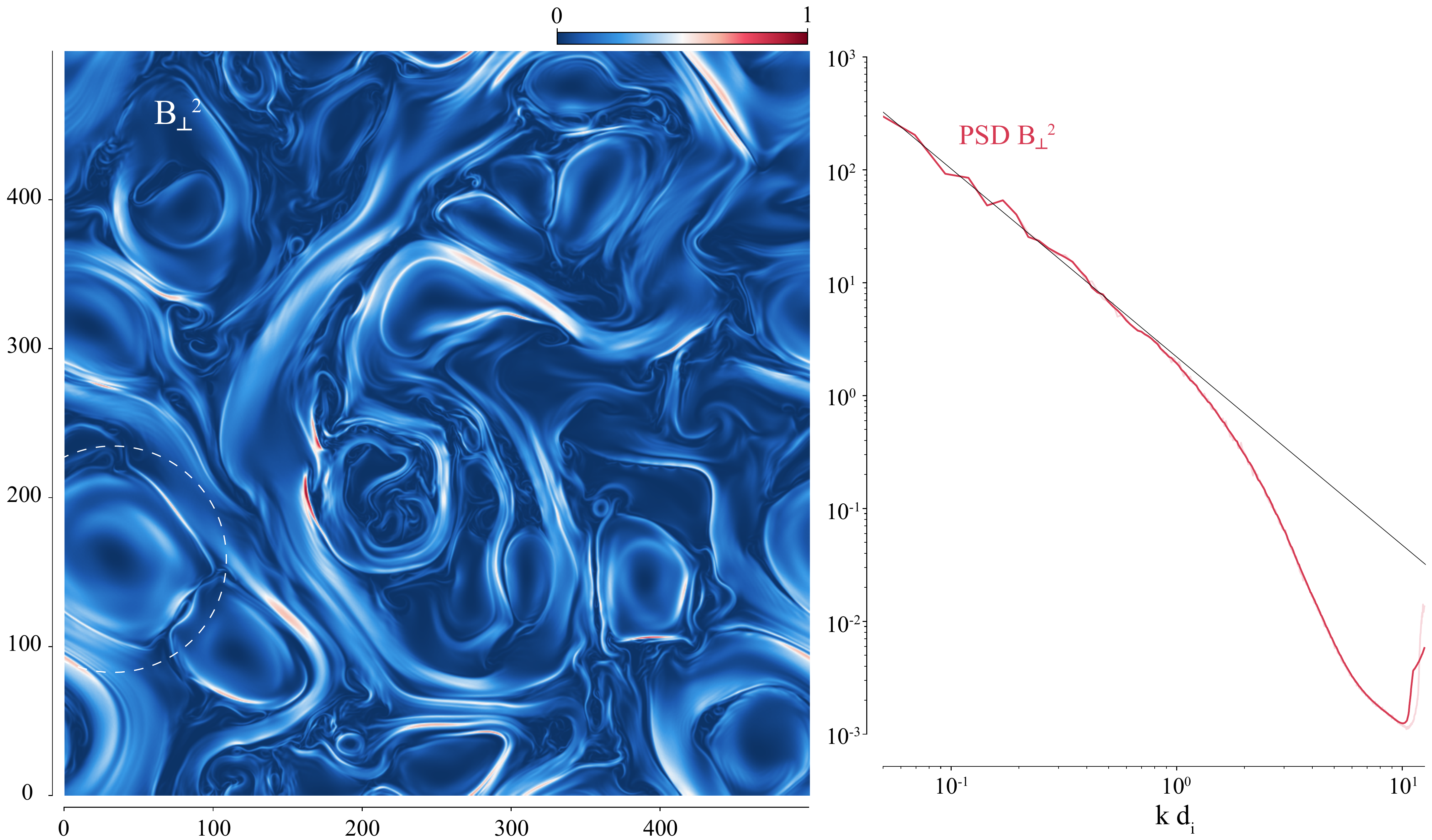}
    \caption{The left-hand panel provides a map of the perpendicular (or in-plane) magnetic field fluctuations squared. The right-hand panel shows their Power Spectral Density, providing a guideline with slope -5/3.}
    \label{fig:spectra}
\end{figure*}

During the first of the two simulation steps, a turbulent plasma is obtained by letting the energy of initial perturbations cascade from large to small scales. Sine modes perturbations are initialised at time 0 in both the in-plane magnetic and velocity fields, on top of a guiding magnetic field $\mathbf{B}_0$ purely out of the simulation plane. All particles are created with velocities following a Maxwellian distribution, using a thermal speed equal to the Alfv\'en speed. At time 500 $\omega_{ci0}^{-1}$ (corresponding to a physical time of 1740~s with the values of Table \ref{tab:decay_param}), the turbulence has developed into the omni-directional Power Spectral Density shown in Figure \ref{fig:spectra}, defined and used in the previous studies of \citet{franci2015apj} and \citet{behar2022angeo}. The spectrum displays a Kolmogorov-like scaling over the inertial, MHD scales, following the black guide line with slope -5/3, to then adopt a much steeper slope at ion kinetic scales, similarly to the results of \citet{franci2015apj} for a very similar simulation. At high spatial frequencies, a flattening of the spectrum is found, corresponding to an energy range in which the noise of the particles is adding up to the cascading energy. At the highest frequencies, we find a sharp increase in energy due to the noise of the finite differences used by the algorithm, a feature shared with the results of \citet{franci2015apj}.

The background values of this run are given in Table \ref{tab:decay_param}. The simulation domain is 500 $d_{i0}$ (corresponding to 65~733~km) large, and a regular grid with 2000 x 2000 nodes is used, corresponding to a grid spacing $\Delta x$ of $0.25 d_{i0}$. The resolved wave vectors are consequently within $[0.0062, 12.4] d_{i0}^{-1}$. The initial perturbations are injected with wave vectors in the range $[0.0062, 0.1] d_{i0}^{-1}$, and only the remaining, non-perturbed spatial scales are shown in Figure \ref{fig:spectra}. This simulation uses 4000 particles per grid node (equivalently cell). A time step $\Delta t = 0.025 \omega_{ci0}^{-1}$ is used during this first step, while a twice finer time resolution $\Delta t = 0.0125 \omega_{ci0}^{-1}$ is needed to properly solve the physics of the tail during Step 2 (cf next sections).

This turbulent plasma has an average out-of-plane magnetic field of precisely $<B_z> = 1$, but an average total magnitude of $<B> = 1.06$. Compared to its laminar analogue, which is defined to be out-of-plane with amplitude 1 everywhere in the domain, we find a magnetic energy density (or magnetic pressure) 12\% larger in the turbulent plasma, due to its additional in-plane component.

\section{General comparison}\label{sec:comp}

\begin{figure*}
    \centering
    \includegraphics[width=\textwidth]{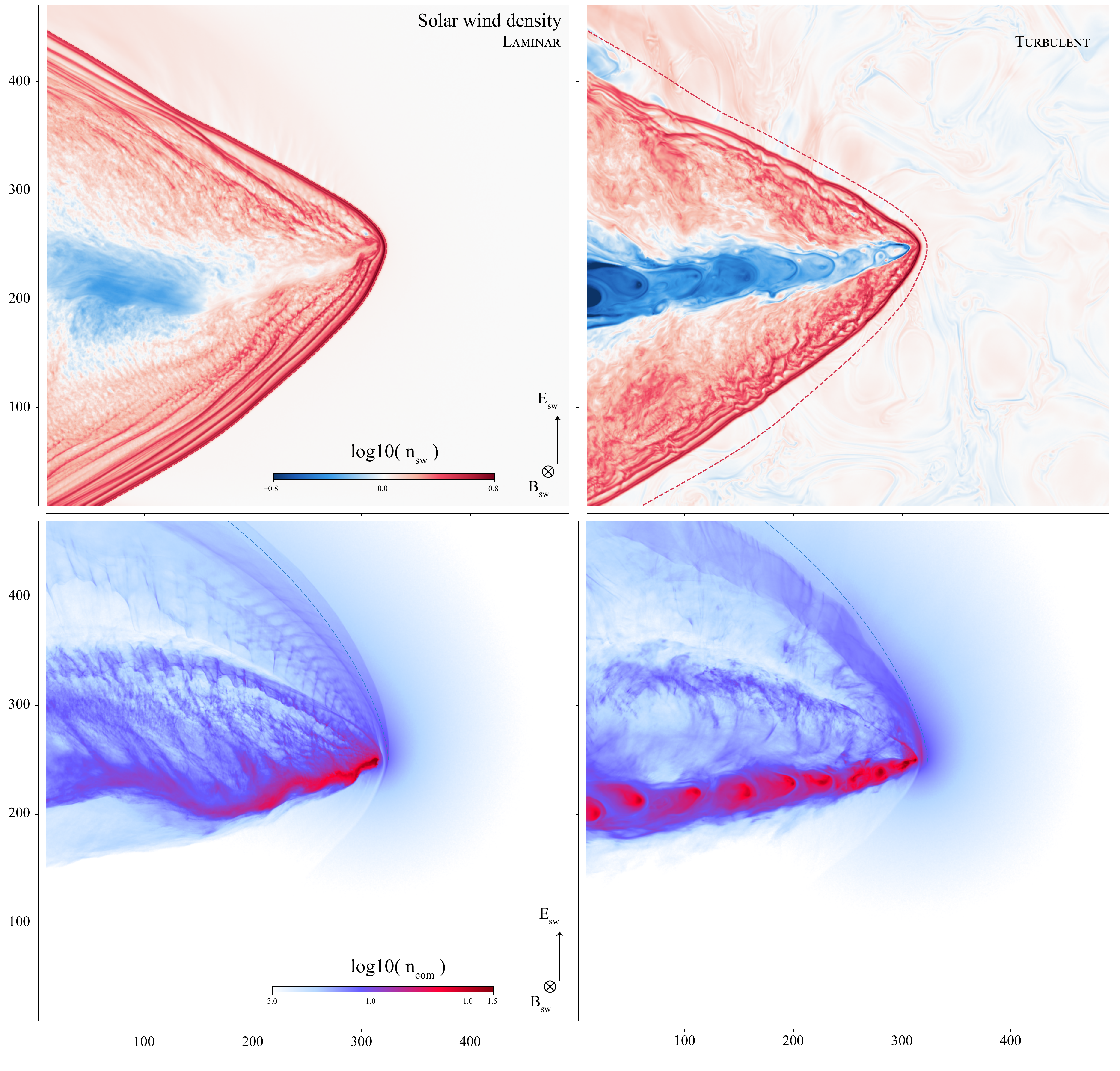}
    \caption{The upper row shows the density of the solar wind ions while the lower row presents the density of the cometary ions. The left column corresponds to laminar upstream conditions, while the right column corresponds to turbulent upstream conditions. The position of the solar wind overdensity in the laminar case is reported in the turbulent case with the dashed red line. Similarly, the estimated gyration of the cometary ions in the plume in the laminar case is shown in the lower row with a blue dashed line.}
    \label{fig:dens}
\end{figure*}

 During the first time interval of Step 2, cometary ions are steadily increasing in number as the magnetosphere develops, before reaching a pseudo steady-state: the total number of particles in the domain evolves around a constant value. After dynamical equilibrium is reached, we simulate further the interaction, for an additional 150 $\omega_{ci0}^{-1}$.

The density of both species are displayed in Figure \ref{fig:dens}, showing one snapshot when dynamical equilibrium is reached, for both the laminar and the turbulent upstream conditions. The solar wind over-density position as well as the pick-up plume, taken in the laminar case (left column), are reported in the turbulent results using dashed lines. We find that both the solar wind over-density and the plume are reduced in size in the turbulent case. On average, the nose of the overdensity is 4 $d_{i0}$ further upstream with laminar upstream conditions. As for the plume, we can estimate the gyration of the ions using their simple upstream gyroradius (the gyroradius a cometary test-particle would have in the upstream wind). For the laminar conditions, with an homogeneous magnetic field, the value is $R_\text{laminar} = 180~d_{i0}$ everywhere in the domain. But in the turbulent case, the magnetic field now has an additional in-plane component, which needs to be accounted for when calculating the gyroradius, which involves the ion velocity component perpendicular to the magnetic field. In addition, the amplitude of the magnetic field is also larger on average in the domain, as described in the previous section. One can compute the local value of the gyroradius, and find an average value over the domain of $R_\text{turbulent} = 147~d_{i0}$, significantly smaller than the laminar value. The lower row of Figure \ref{fig:dens} uses a cycloid of radius 180, shown with a blue dashed line, showing a good match with the laminar plume, while the turbulent plume is found to be smaller.

Whether we look at this interaction from a kinetic point of view \citep{behar2018aa}, involving these gyration scales, or a fluid point of view, involving the upstream magnetic pressure, we find in both cases that the plasma structures around such a comet are expected to be smaller in the turbulent case, which we indeed verified with the present simulations. It should be noted however that the definition of the laminar plasma is a choice, arbitrary to some extent, and one may argue that it could very well be defined in such a way that the magnetic energy density or the gyroradius are equal in both the laminar and the turbulent case.

It should be pointed out that the laminar interaction already shows some level of complexity, with some obvious wave patterns within the magnetosphere in the upper part of the cometary ion density (likely similar to the bi-ion acoustic waves also found by \citet{bagdonat2002emp}, but also in the lower part of the solar wind density (see also the work of \citet{koenders2016aa} on low frequency waves at comet 67P/CG). These fluctuations and their fate when considering a turbulent upstream input, as well as their potential contribution to the inner-magnetosphere turbulence, is yet another important future direction to explore.

The Hall escape channel is shown stemming from the comet's inner region in the lower panels of Figure~\ref{fig:dens}. In the laminar case (bottom left panel), the homogeneous incoming magnetised solar wind results in a continuous cometary ion acceleration and in turn in a continuous cometary ion structure, forming the Hall escape channel. On the contrary, in the turbulent case (bottom right panel), the turbulent nature of the incoming magnetised solar wind is responsible for the generation of discontinuities in this same Hall escape channel, which is now found to be composed of discrete, high density cometary ion bubbles, of similar density as the inner coma.

\begin{figure*}
    \centering
    \includegraphics[width=.8\textwidth]{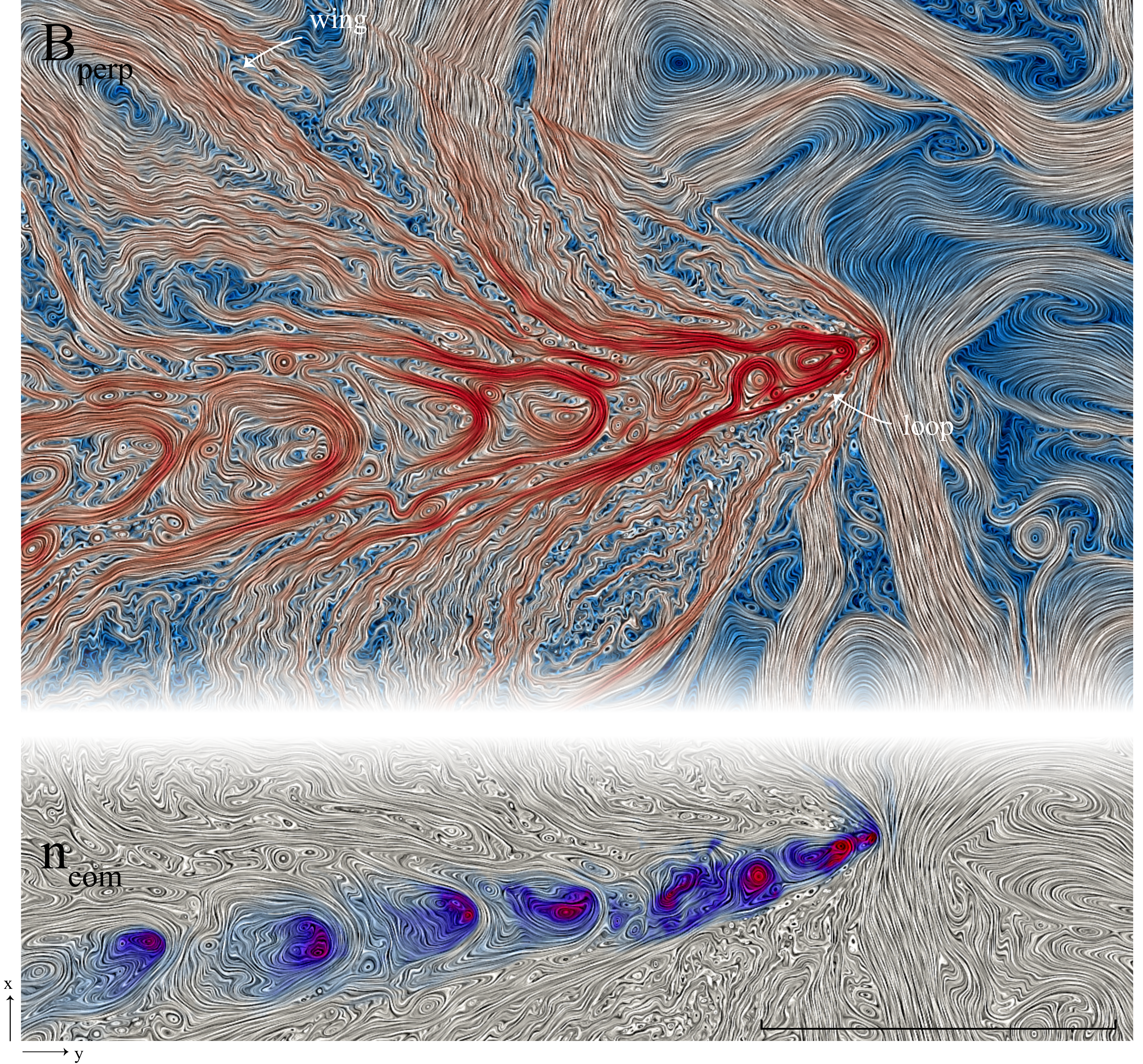}
    \caption{The upper panel provides a Line Integral Convolution of the in-plane magnetic field component, coupled to its colour-coded amplitude. The lower panel shows the same LIC representation, coupled to the cometary ion density along the comet tail. In the lower panel, lower-right corner, the horizontal line length corresponds to 100 $d_i$.}
    \label{fig:lic}
\end{figure*}

The modifications of the solar wind turbulent flow are highlighted in Figure~\ref{fig:lic}, which shows the in-plane magnetic field lines, using a Line Integral Convolution (LIC) representation \citep{loring2015}. In the upper panel, the in-plane magnetic field lines are coloured using the amplitude of the same in-plane field, while the lower panel uses the density of cometary ions superposed to the LIC representation. Cometary ion bubbles are found to be enclosed in magnetic islands, downstream of the nucleus, islands which were not present in the upstream turbulent wind. We identify two main regions in this magnetosphere. First the ``wings'', in which deformed upstream perpendicular field structures can be recognised. And second, the comet tail in which most of the cometary ions are confined, corresponding to very low solar wind densities in Figure~\ref{fig:dens}. There, newly formed magnetic field islands of various sizes can be observed, and dubbed as ``loops'' on Figure~\ref{fig:lic}. In the next two sections, we explore the origin of these two regions. 

\section{Disconnection of the comet's head}

The pile-up and draping of the magnetic field, which so far in the literature was discussed in the context of a laminar interaction, corresponding to an homogeneous upstream magnetic field, in turn affect the turbulent structures of the solar wind magnetic field as well. To our knowledge, the consequences of cometary-induced pile-up and draping of the magnetic field on the incoming turbulent, heterogeneous plasma flow have not been considered or investigated so far.

\begin{figure}
    \centering
    \includegraphics[width=.85\linewidth]{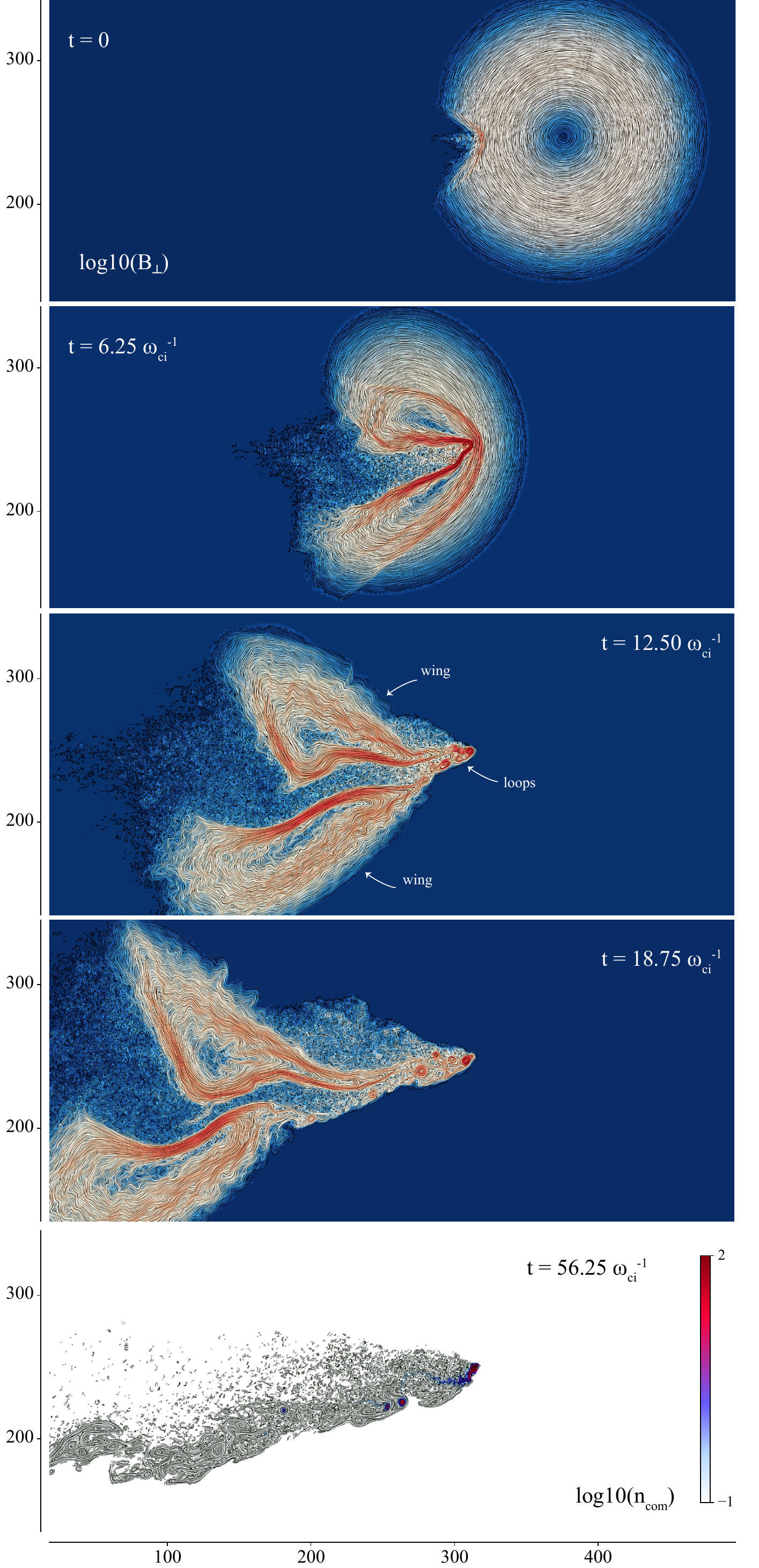}
    \caption{Each panel provides a snapshot of the numerical experiment. All panels use a LIC representation of the in-plane magnetic field lines. The first four rows use the amplitude of the in-plane magnetic field for the colours, while the last panel uses the cometary ion density.}
    \label{fig:ersatz_lic}
\end{figure}

To illustrate the mechanism at the origin of the formation of these high density bubbles and the complex magnetic field structures within the comet's tail, we have designed a dedicated numerical experiment to isolate the interaction between a single upstream perpendicular magnetic field structure and the coma, in an otherwise laminar upstream flow. The laminar version of the interaction is resumed from its steady state shown in Figure \ref{fig:dens}, and an ersatz of a magnetic island (i.e. a flux rope in the 3-dimensional case) is introduced upstream of the comet. Eventually, the comet meets and interacts with this artificial structure, just as it does with structures of the fully turbulent flow.

The magnetic field island ersatz is defined by adding a perpendicular component to the otherwise homogeneous and out-of-plane background magnetic field. This additional perpendicular component is characterised by circular in-plane field lines, and an amplitude depending on the distance to the centre of these circular lines, with a maximum value at some distance from the centre. We have chosen a Gaussian profile with a maximum value of 0.5 the background magnetic field at 50 $d_{i0}$ from the centre of the structure, and a standard deviation of 25 $d_{i0}$. This artificial structure is meant to mimic the large scale vortices found in the perpendicular magnetic field shown in the left hand-side of Figure \ref{fig:spectra}, with diameters about 50 to 100 $d_{i0}$, with a larger amplitude found away from their centres. Such a structure is indicated by a dashed circle in the left panel of Figure \ref{fig:spectra}. This ersatz is not meant to be a perfect replica of such structures however: no further considerations are taken into account other than a divergence-free, circular structure of about 100 $d_{i0}$ in diameter, with a realistic maximum found at some distance from its centre.

This ersatz can be seen still undisturbed upstream of the comet in the uppermost panel of Figure \ref{fig:ersatz_lic}, defining the relative time $t=0$ for this experiment. The figure uses the same LIC representation as Figure \ref{fig:lic}, with a threshold of 0.01 on the in-plane (perpendicular) magnetic field: if the perpendicular magnetic field is smaller than 0.01, no LIC is shown. The background colourmaps of the first four panels from the top, showing four successive and equally separated times of the experiment, additionally give the amplitude of the perpendicular magnetic field, while for the final time of the experiment, given in the bottom panel, the density of cometary ions is used.

The time $t=0$ is chosen as the magnetic island is just starting to interact with the dense inner coma. As a result, the left-most field lines are found to be deformed, draped inward the island, with a corresponding increase of the field amplitude -- pile-up -- seen with red tones at the left-most boarder of the structure. As the magnetic field amplitude increases and the length scale between the anti-parallel magnetic field lines shortens, a strong current sheets forms around the comet nucleus location (second panel). This strong current sheet is likely unstable to the tearing instability, eventually forming shorter scale magnetic islands (third panel).  

This series of snapshots gives us a very tangible representation of how the comet pierces through an upstream magnetic field structure, similar to a projectile through an obstacle. After ``impact'', we find two remains of the initial structure with high perpendicular magnetic field amplitudes looping around two distinct centres, on each side of the comet's head. Within these two wings, the upstream information is conserved to some extent, with field structures intensified and deformed. However, closer to the comet's head and downstream of it, new, smaller structures of even higher perpendicular magnetic field amplitude are produced, seen as smaller scale closed magnetic field loops. Whereas the two wings are found to be more or less static in the solar wind reference frame, just as their parent upstream structure, these smaller scale intense loops have a speed closer to the comet itself. Or equivalently, described in the object reference frame, whereas the wings are advected downstream at more or less the speed of the solar wind, the new loops are transported downstream at a much lower speed.

The difference between these two types of structures originate from the density the upstream parent structure meets during its interaction with the coma. On the one hand, the sides, or wings, of the parent island interact with the comet in regions where the density is still dominated by solar wind protons in terms of number density, and the two wings keep frozen into the solar wind dominated flow, conserving some information from upstream. On the other hand, at the centre of the interaction region, the plasma is largely dominated by the cometary ions. The upstream magnetic field, now with the additional perpendicular component, is piling up in this drastically reduced plasma mean velocity, reaching higher amplitudes than in the wings. And most interestingly, this rising magnetic tension, through the Hall electric field, eventually pulls off the dense inner coma, resulting in one main and two secondary high density bubbles found downstream of the comet's head at time 56.25 (bottom panel), enclosed in the newly formed, intense magnetic field loops. Note that at this time, taken much later than the first four snapshots, the wings, not carrying any significant cometary ion density, are by then long gone downstream of the comet. After this build-up and release phenomenon, and in the absence of additional heterogeneous perpendicular magnetic field structures, the inner coma resumes its laminar, continuous Hall escape.

We will now have a closer look at this disconnection event, zooming inside the inner interaction region, and focusing on times around $t=6.25$.

\section{Disconnection at the ion scale}

\begin{figure*}
    \centering
    \includegraphics[width=\linewidth]{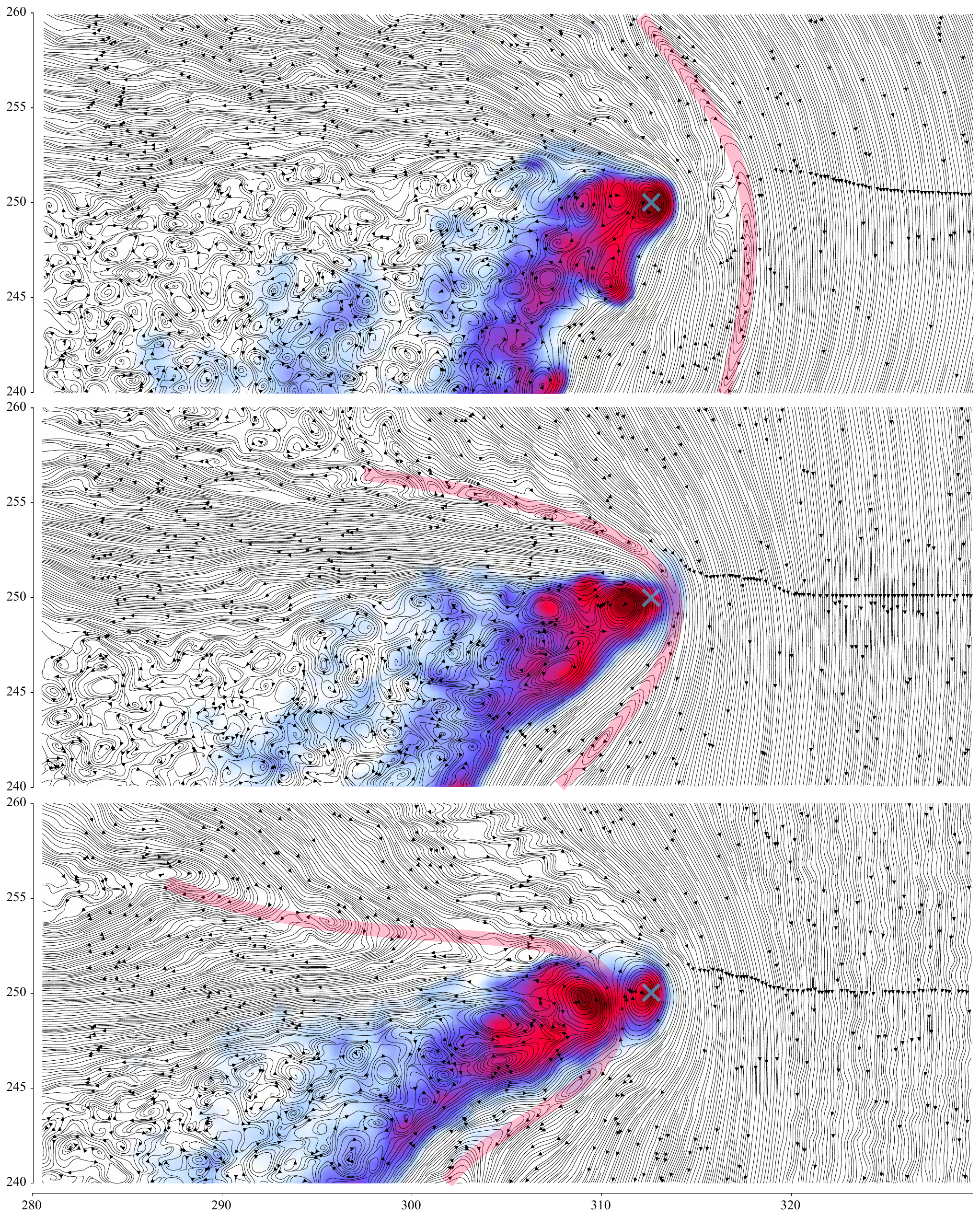}
    \caption{In-plane magnetic field lines, oriented, overlapped to the cometary ion density, taken during the numerical experiment and centre around time $t=6.25$ of Figure \ref{fig:ersatz_lic}. The red band is manually added to visually follow the main vertical axis of the upstream structure. The blue crosses indicate the centre of the comet.}
    \label{fig:ersatz_lines}
\end{figure*}

Figure \ref{fig:ersatz_lines} shows the comet's head during the same experiment, zooming a bit closer in the spatial domain than the previous representation, using a more classic, oriented field line representation superposed to the cometary ion density. These three snapshots are also focused around the time when the centre of the upstream magnetic island passes through the inner coma, corresponding to the second panel of Figure \ref{fig:ersatz_lic}. In the upper panel, the magnetic island can be identified, already significantly piled-up, i.e. strongly compressed along the $x$-direction. As the plasma is faster above and below the inner coma (red tones of the colourmap), the island is additionally draped, and with the upper and lower parts of the island advected downstream much faster than the central part, an elongation of the structure is also seen. Field lines which were initially circular are now found to describe highly eccentric ellipse-like figures, along a main axis highlighted with a manually added red band, describing a bow. Because of this combined compression and elongation, along the red band, we find a line separating field lines parallel and of opposite sense. As the comet continues piercing through the structure, the red band gets significantly draped. In the second panel, multiple reconnection points along the red band have occurred, resulting in new, small scale magnetic islands. At the very tip of the band, where the field keeps piling up, we find one main magnetic reconnection site on which anti-parallel field lines, perpendicular to the flow, keep being compressed against each other.

Under this magnetic tension, the inner ionised coma is slowing down in the plasma frame, or equivalently pushed downstream in the object frame. This can be easily spotted by looking at the blue crosses in each panel, which indicate the centre of the comet. Eventually, the tension becomes strong enough to significantly accelerate the inner coma, which starts moving downstream of the comets centre (in the comet's frame). As soon as this pair left the central region, the continuous ionisation of neutral molecules starts refilling the inner coma, with most of the magnetic tension released. In the last panel of Figure \ref{fig:ersatz_lines}, one can see the newly detached high density cloud enclosed in closed field lines, while upstream of it a fresh inner coma is getting denser, though still much less dense than the detached cometary plasma bubble.

\begin{figure*}
    \centering
    \includegraphics[width=\linewidth]{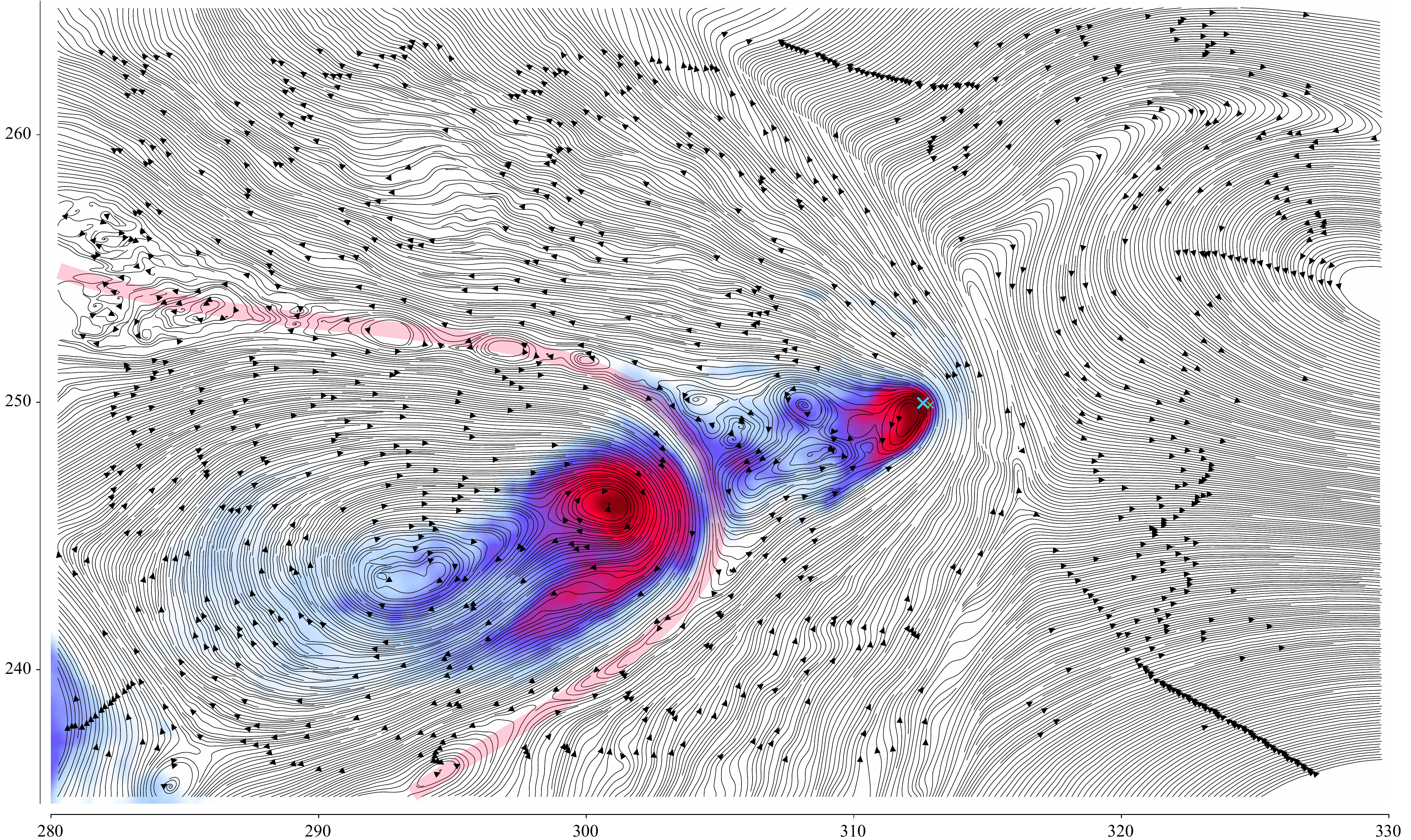}
    \caption{In-plane magnetic field lines, oriented, overlapped to the cometary ion density, taken during the fully turbulent simulation of the interaction. The manually added red band indicates a similar reconnecting current sheet as in Figure \ref{fig:ersatz_lines}. The larger blue cross indicates the centre of the comet, while the smaller cross shows the position of the synthetic probe used in Section \ref{sec:draining}}
    \label{fig:lines}
\end{figure*}

The exact same situation can be spotted in the fully turbulent interaction, with Figure \ref{fig:lines} showing one snapshot when a dense cometary ion cloud just got detached from the central region of the comet, enclosed in a magnetic island, while a series of smaller scale islands (emphasized in the figure by the manually added red band for visualisation purposes) are spotted, not capturing high densities of cometary ions. Despite not designing the upstream magnetic island to be a perfect replica of some structures of the turbulent flow, we were able to reproduce the density disconnection event in great details, illustrating how the pile-up and draping of upstream perpendicular magnetic field structures can disrupt the Hall escape channel of the comet, producing cometary ion bubbles within the tail, together with new, shorter scale, magnetic field structures not present upstream. Whereas our ersatz was perfectly circular, the mechanism happens in the same manner with more complex field structures, which do not even need to present closed loops, as long as upstream anti-parallel magnetic field lines pile-up in front of the inner coma to generate strong, unstable, current sheets in the exact same way, resulting in the global picture of Figures \ref{fig:dens} and \ref{fig:lic}.

\section{Draining of the inner coma}\label{sec:draining}

\begin{figure*}
    \centering
    \includegraphics[width=\linewidth]{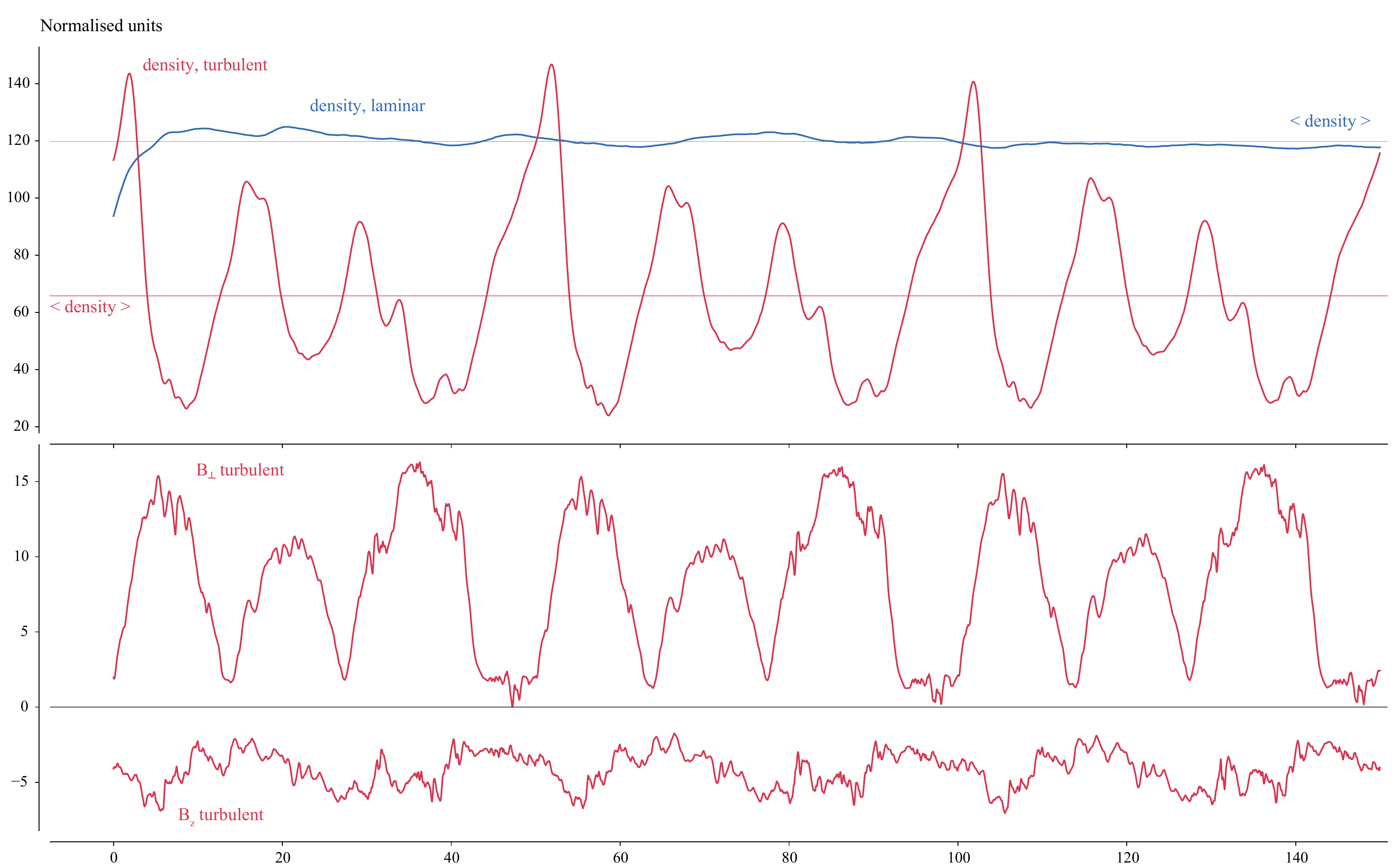}
    \caption{Time series of the total density (upper panel) and the in-plane and out-of-plane magnetic field components (lower panel), taken in the inner coma, given in normalised units. The horizontal lines of the upper panel correspond to averaged values of the time series.}
    \label{fig:time_series}
\end{figure*}

This tension/release phenomenon in a fully turbulent flow leads to a pseudo-periodic creation of cometary ion bubbles. The effect of this dynamical flushing is most spectacular when following through time the plasma parameters right within the inner coma, i.e. close to the nucleus. A synthetic probe was used within the simulation, static in the comet frame, with a position shown in Figure \ref{fig:lines} (a smaller cross almost superposed to the centre of the comet). At the probe's position, all plasma parameters (fields values and particles moments) were recorded, for a duration of three full periodic injection of the upstream turbulent flow. Figure \ref{fig:time_series} provides the time series of the ion density in the upper panel, comparing the turbulent and laminar runs, while the lower panel gives the perpendicular and out-of-plane magnetic field for the turbulent case.

The phenomenon previously illustrated in the spatial domain can also be recognised in these time series: each sudden drop of density (upper red line) corresponding to the time when a bubble is disconnected from the inner coma and moves downstream, a maximum in the perpendicular magnetic field (lower panel) follows shortly after. Overall, we observe that the two quantities are phased of about $\pi/2$, with maxima (respectively minima) of the magnetic field found at a decreasing (respectively increasing) density, crossing its average value. This phasing results from the competing pressures (magnetic on one hand, dynamic and thermal on the on other hand), never reaching an equilibrium in this dynamic inner region. During this pseudo-periodic draining of the inner coma, the density reaches maximum values 147 times higher than the upstream mean density, to then drop all the way down to values around 24. During these drops, the perpendicular magnetic field reaches values of 16 times the background magnetic field, with minimum values around 2, and locally even reaching 0. The out-of-plane magnetic field $B_z$ is found to vary much less, around an average value of -4.

In these time series recorded during the turbulent interaction, the periodicity corresponding to the cyclic turbulent flow injection is clearly seen. We can deduce from the upper panel density curves that one full injection of the turbulent plasma eventually leads to the formation of three main density bubbles.

The inner coma density recorded at the same point during the laminar interaction shows a much more stable behaviour, around an average value of 120 times the upstream density. The average value in the turbulent case is only 55\% of that. Reformulating this number, in the limit of the model exposed above, we find an inner coma on average almost twice thinner when we add turbulent fluctuations to the upstream flow, with variations from 20\% to 122\%, compared to a laminar case with all background parameters equal.

\section{Summary}

This work demonstrates how the global picture of the interaction between an object and a turbulent solar wind, as provided by the dedicated code \emph{Menura}, can be approached and analysed, for the first time. On this first application, we have focused on the macroscopic effects of the upstream turbulence on the magnetosphere, especially studying the apparition of high density cometary ion bubbles within the comet tail, not present in the laminar case.

Using an isolated ersatz of a large scale, upstream magnetic field structure, we have reproduced the two main characteristics found in the fully turbulent interaction. This illustrates that it is the large scale turbulent structures, carrying most of the turbulent energy, which may impact the most the magnetosphere. Another dedicated parameter study will be necessary to explore the topic further, using various initialisation of the turbulence. Based on these early results however, an important aspect of such a future study would be the role of the amplitude of the initial fluctuations, as well as their range (for instance including even larger $k$ values, i.e. a larger simulation domain).

First, we see the creation of two different regions in the induced cometary magnetosphere, the wings, in which the plasma flow is disturbed to some extent, but keeps some of its upstream information and relative speed, and the tail, in which newly formed structures move downstream much slower, showing much more intense magnetic field. Second, we were able with this single structure to reproduce the disconnection of the the comet's head in terms of cometary ion density, leading to high density bubbles correlating to the new magnetic field structures within the tail, just as observed in the fully turbulent run. 

The creation of these cometary ion bubbles goes hand in hand with the pseudo-periodic draining of the inner-most coma, within which very large fluctuations in terms of density and electric and magnetic field amplitudes are found. These variations in terms of density are ranging within 20\% and 122\% of the average laminar density. Another striking result is that on average, the turbulent interaction results in a dramatically lower average density in the inner coma, only 55\% of that of the laminar interaction.

The temporal evolution of the overdensity position is greater in the turbulent case (not shown in this publication). We could not yet find the right combination of parameters which could account for this precise evolution. The problem is indeed fairly complex, as both the upstream parameters and the inner-coma parameters are highly dynamic, involving different time scales.\\

The implications of such a reduced average density of the inner coma would be profound for all aspects of its physics. A first example is the balance between the various pressures, magnetic, ram, and thermal, each depending on the magnetic field and the plasma density. Changing so significantly the pressures, within the magnetosphere as well as upstream of it, will necessarily and significantly move the various plasma boundaries from laminar to turbulent upstream conditions. A dynamic input is also expected to introduce more dynamics along these boundaries. Another major topic concerned by this redefined plasma density is the physico-chemistry of the inner coma, with charge-exchange processes heavily based on the ion and electron density.

This first work paves the road to a comparative planetary science study at other types of bodies, such as ionospheres (Mars, Venus) and permanent dipoles (Earth, Mercury). The potential effect of upstream turbulence on the escape of planetary ions may introduce a new argument in the topic of the atmosphere evolution.

Though this 2-dimensional approach would benefit from a future back up by additional 3-dimensional runs, the basic phenomenon illustrated in this publication should remain, to some extent, valid: upstream perpendicular magnetic field structures will still pile-up and drape, self-consistently forming strong, unstable current sheets, at the location where the cometary plasma is densest. Whether the effect on the planetary ions will be that spectacular is, however, not obvious. The nature of the upstream turbulence, and especially its large scale structures relevant for this study, are also quite different when solving the problem with a third dimension; see for instance the study of \citep{franci2018apj}, highlighting some differences between 2D and 3D cascades at MHD scales.

Another important direction to explore is the opposite problem: the effect of the magnetosphere on the solar wind turbulence. This is a somewhat more delicate question, which might necessitate a more refined approach.

\begin{acknowledgements}

E. Behar acknowledges support from Swedish National Research Council, Grant 2019-06289. This work was granted access to the HPC resources of IDRIS under the allocation 2021-AP010412309 made by GENCI. Work at LPC2E and Lagrange was partly funded by CNES.

\end{acknowledgements}

%
%

\bibliographystyle{aa}
\bibliography{eb_biblio} 

\begin{thebibliography}{36}
\expandafter\ifx\csname natexlab\endcsname\relax\def\natexlab#1{#1}\fi

\bibitem[{Alfven(1957)}]{alfven1957theory}
Alfven, H.~t. 1957, Tellus, 9, 92

\bibitem[{Antonova \& Stepanova(2021)}]{antonova2021fass}
Antonova, E.~E. \& Stepanova, M.~V. 2021, Frontiers in Astronomy and Space
  Sciences, 8

\bibitem[{Bagdonat \& Motschmann(2002)}]{bagdonat2002emp}
Bagdonat, T. \& Motschmann, U. 2002, Earth, Moon, and Planets, 90, 305

\bibitem[{Behar {et~al.}(2022)Behar, Fatemi, Henri, \&
  Holmstr\"om}]{behar2022angeo}
Behar, E., Fatemi, S., Henri, P., \& Holmstr\"om, M. 2022, Annales Geophysicae
  Discussions, 2022, 1

\bibitem[{Behar {et~al.}(2018)Behar, Tabone, Saillenfest, Henri, Deca,
  Lindkvist, Holmstr{\"o}m, \& Nilsson}]{behar2018aa}
Behar, E., Tabone, B., Saillenfest, M., {et~al.} 2018, Astronomy \&
  Astrophysics, 620, A35

\bibitem[{{Behar, E.} {et~al.}(2018){Behar, E.}, {Nilsson, H.}, {Henri, P.},
  {Bercic, L.}, {Nicolaou, G.}, {Stenberg Wieser, G.}, {Wieser, M.}, {Tabone,
  B.}, {Saillenfest, M.}, \& {Goetz, C.}}]{behar2018aa_b}
{Behar, E.}, {Nilsson, H.}, {Henri, P.}, {et~al.} 2018, A\&A, 616, A21

\bibitem[{{Biermann}(1951)}]{biermann1951}
{Biermann}, L. 1951, \zap, 29, 274

\bibitem[{Bruno \& Carbone(2005)}]{bruno2005lrsp}
Bruno, R. \& Carbone, V. 2005, Living Reviews in Solar Physics, 2, 4

\bibitem[{{Cheng} {et~al.}(2022){Cheng}, {Wang}, \& {Li}}]{cheng2022apj}
{Cheng}, L., {Wang}, Y., \& {Li}, X. 2022, \apj, 928, 121

\bibitem[{Cranmer {et~al.}(2015)Cranmer, Asgari-Targhi, Miralles, Raymond,
  Strachan, Tian, \& Woolsey}]{cranmerptrs2015}
Cranmer, S.~R., Asgari-Targhi, M., Miralles, M.~P., {et~al.} 2015,
  Philosophical Transactions of the Royal Society A: Mathematical, Physical and
  Engineering Sciences, 373, 20140148

\bibitem[{DeForest {et~al.}(2015)DeForest, Matthaeus, Howard, \&
  Rice}]{deforest2015pj}
DeForest, C.~E., Matthaeus, W.~H., Howard, T.~A., \& Rice, D.~R. 2015, The
  Astrophysical Journal, 812, 108

\bibitem[{Dong {et~al.}(2015)Dong, Fang, Brain, McFadden, Halekas, Connerney,
  Curry, Harada, Luhmann, \& Jakosky}]{dong2015grl}
Dong, Y., Fang, X., Brain, D.~A., {et~al.} 2015, Geophysical Research Letters,
  42, 8942

\bibitem[{Dungey(1961)}]{dungeyprl1961}
Dungey, J.~W. 1961, Phys. Rev. Lett., 6, 47

\bibitem[{D’Amicis {et~al.}(2020)D’Amicis, Telloni, \&
  Bruno}]{damicis2020fass}
D’Amicis, R., Telloni, D., \& Bruno, R. 2020, Frontiers in Physics, 8

\bibitem[{El-Alaoui {et~al.}(2021)El-Alaoui, Walker, Weygand, Lapenta, \&
  Goldstein}]{elalaoui2021fass}
El-Alaoui, M., Walker, R.~J., Weygand, J.~M., Lapenta, G., \& Goldstein, M.~L.
  2021, Frontiers in Astronomy and Space Sciences, 8

\bibitem[{Franci {et~al.}(2015)Franci, Landi, Matteini, Verdini, \&
  Hellinger}]{franci2015apj}
Franci, L., Landi, S., Matteini, L., Verdini, A., \& Hellinger, P. 2015, The
  Astrophysical Journal, 812, 21

\bibitem[{Franci {et~al.}(2018)Franci, Landi, Verdini, Matteini, \&
  Hellinger}]{franci2018apj}
Franci, L., Landi, S., Verdini, A., Matteini, L., \& Hellinger, P. 2018, The
  Astrophysical Journal, 853, 26

\bibitem[{Goetz(2022)}]{goetz2022ssr}
Goetz, C. 2022, SSR

\bibitem[{Guio \& Pécseli(2021)}]{guio2021fass}
Guio, P. \& Pécseli, H.~L. 2021, Frontiers in Astronomy and Space Sciences, 7,
  107

\bibitem[{Hansen {et~al.}(2007)Hansen, Bagdonat, Motschmann, Alexander, Combi,
  Cravens, Gombosi, Jia, \& Robertson}]{hansen2007ssr}
Hansen, K.~C., Bagdonat, T., Motschmann, U., {et~al.} 2007, Space Science
  Reviews, 128, 133

\bibitem[{Kallio {et~al.}(2012)Kallio, Chaufray, Modolo, Snowden, \&
  Winglee}]{kallio2012}
Kallio, E., Chaufray, J.-Y., Modolo, R., Snowden, D., \& Winglee, R. 2012,
  Modeling of Venus, Mars, and Titan, ed. K.~Szego (New York, NY: Springer US),
  267--307

\bibitem[{{Koenders, C.} {et~al.}(2016){Koenders, C.}, {Perschke, C.}, {Goetz,
  C.}, {Richter, I.}, {Motschmann, U.}, \& {Glassmeier, K.
  H.}}]{koenders2016aa}
{Koenders, C.}, {Perschke, C.}, {Goetz, C.}, {et~al.} 2016, A\&A, 594, A66

\bibitem[{{Kolmogorov}(1941)}]{kolmogorov1941}
{Kolmogorov}, A. 1941, Akademiia Nauk SSSR Doklady, 30, 301

\bibitem[{Lakka {et~al.}(2019)Lakka, Pulkkinen, Dimmock, Kilpua, Ala-Lahti,
  Honkonen, Palmroth, \& Raukunen}]{lakka2019angeo}
Lakka, A., Pulkkinen, T.~I., Dimmock, A.~P., {et~al.} 2019, Annales
  Geophysicae, 37, 561

\bibitem[{Lakka {et~al.}(2017)Lakka, Pulkkinen, Dimmock, Osmane, Honkonen,
  Palmroth, \& Janhunen}]{lakka2017angeo}
Lakka, A., Pulkkinen, T.~I., Dimmock, A.~P., {et~al.} 2017, Annales
  Geophysicae, 35, 907

\bibitem[{{Loring} {et~al.}(2015){Loring}, {Karimabadi}, \&
  {Rortershteyn}}]{loring2015}
{Loring}, B., {Karimabadi}, H., \& {Rortershteyn}, V. 2015, in Astronomical
  Society of the Pacific Conference Series, Vol. 498, Numerical Modeling of
  Space Plasma Flows ASTRONUM-2014, ed. N.~V. {Pogorelov}, E.~{Audit}, \& G.~P.
  {Zank}, 231

\bibitem[{Ma {et~al.}(2008)Ma, Altwegg, Breus, Combi, Cravens, Kallio, Ledvina,
  Luhmann, Miller, Nagy, Ridley, \& Strobel}]{Ma2008}
Ma, Y.-J., Altwegg, K., Breus, T., {et~al.} 2008, Space Science Reviews, 139,
  311

\bibitem[{Moissard {et~al.}(2022)Moissard, Savoini, Fontaine, \&
  Modolo}]{moissard2022essoar}
Moissard, C., Savoini, P., Fontaine, D., \& Modolo, R. 2022, Earth and Space
  Science Open Archive, 21

\bibitem[{{Niedner} \& {Brandt}(1978)}]{niedner1978apj}
{Niedner}, M.~B., J. \& {Brandt}, J.~C. 1978, \apj, 223, 655

\bibitem[{{Parker}(1958)}]{parker1958apj}
{Parker}, E.~N. 1958, \apj, 128, 664

\bibitem[{Rakhmanova {et~al.}(2021)Rakhmanova, Riazantseva, \&
  Zastenker}]{rakhmanova2021fass}
Rakhmanova, L., Riazantseva, M., \& Zastenker, G. 2021, Frontiers in Astronomy
  and Space Sciences, 7, 115

\bibitem[{Romanelli {et~al.}(2019)Romanelli, DiBraccio, Modolo, Leblanc,
  Espley, Gruesbeck, Halekas, Mcfadden, \& Jakosky}]{romanelli2019grl}
Romanelli, N., DiBraccio, G., Modolo, R., {et~al.} 2019, Geophysical Research
  Letters, 46, 10977

\bibitem[{Ruhunusiri {et~al.}(2020)Ruhunusiri, Howes, \&
  Halekas}]{ruhunusiri2020jgr}
Ruhunusiri, S., Howes, G.~G., \& Halekas, J.~S. 2020, Journal of Geophysical
  Research: Space Physics, 125, e2020JA028100, e2020JA028100
  10.1029/2020JA028100

\bibitem[{Saur(2021)}]{saur2021fass}
Saur, J. 2021, Frontiers in Astronomy and Space Sciences, 8, 56

\bibitem[{Schunk \& Nagy(2009)}]{schunk_nagy_2009}
Schunk, R. \& Nagy, A. 2009, Ionospheres: Physics, Plasma Physics, and
  Chemistry, 2nd edn., Cambridge Atmospheric and Space Science Series
  (Cambridge University Press)

\bibitem[{Vourlidas {et~al.}(2007)Vourlidas, Davis, Eyles, Crothers, Harrison,
  Howard, Moses, \& Socker}]{vourlidas2007apj}
Vourlidas, A., Davis, C.~J., Eyles, C.~J., {et~al.} 2007, The Astrophysical
  Journal, 668, L79

\end{thebibliography}

\end{document}